\documentclass[referee]{raa}            

\usepackage{graphicx,times}             
\usepackage{natbib}
\usepackage{amssymb,amsmath}
\usepackage[flushleft]{threeparttable}
\bibpunct{(}{)}{;}{a}{}{,}

\usepackage[a4paper=true,driverfallback=dvipdfm,pagebackref=true]{hyperref}
\hypersetup{colorlinks = true, linkcolor = green, anchorcolor = red, citecolor = blue, filecolor = red, pagecolor = red, urlcolor = red}

\begin{document}

   \title{Light Curves Analysis and Period Study of Two Eclipsing Binaries UZ Lyr and BR Cyg}

   \volnopage{Vol.0 (20xx) No.0, 000--000}      
   \setcounter{page}{1}          

   \author{K. Y. Roobiat
      \inst{1}
   \and R. Pazhouhesh
      \inst{1}
  }

   \institute{Department of Physics, Faculty of Sciences, University of Birjand, Birjand, Iran.{\it rpazhouhesh@birjand.ac.ir}\\
\vs\no
   {\small Received~~20xx month day; accepted~~20xx~~month day}}

\abstract{Two eclipsing binary systems UZ Lyr and BR Cyg are the semi-detached types whose secondary component fill its Roche lobe. Although radial velocity and light curves of these systems have already been investigated separately, both radial velocity and light curves of them are analyzed simultaneously for the first time in the present study . Also, the orbital period changes of these systems are studied. Our results show that the mass transfer between components have negligible effects on the orbital period changes of these systems, but two light-time effects are the reasons of the periodic behavior of the O-C curve for UZ Lyr. We could not remark more information about orbital period changes for BR Cyg, but we find a new orbital period for it. By radial velocity and light curves analysis we find a clod spot on the secondary components of BR Cyg. The new geometrical and physical parameters of both systems are obtained and their positions on H-R diagram demonstrated.
}

   \maketitle

%
\section{Introduction}           

The UZ Lyr has been considered for a long time as an Algol-type eclipsing binary system (\citealt{Nijland1931}). However, this system has not been photometry periodically, and a detailed analysis has not been available for its light curve. In this regard, \cite{Koch1979} classified this system as "forgotten". This system has been included in the catalog of variable stars in the Kepler project In 2009 (\citealt{Pigulski2009}), and it has been comprised of the first Kepler eclipsing binary catalog, with a specified temperature ratio of $\frac{T_2}{T_1}=0.51793$ and an effective temperature of $T_eff=11061 \; K$ in 2011 (\citealt{Prsa2011}). This temperature ratio has been changed to $0.51276$ in the second Kepler eclipsing binary catalog (\citealt{Slawson2011}). By combining data from the Kepler project with those from HESS, KIS, and 2MASS projects, the temperature of the primary and secondary components of the system has been reported $15058 \pm 515\; K$ and $10411 \pm 2484 \; K$, respectively by \cite{Armstrong2013}.\\
The radial velocity curve of UZ Lyr has been obtained by \cite{Matson2017}, and the value of $q =\frac{m_2}{m_1}= 0.23 \pm 0.01$ has been reported as the spectroscopic mass ratio. The absolute masses of the primary and secondary components have been calculated $4.05 \pm 0.3 \; M_{\odot}$ and $0.92 \pm 0.07 \; M_{\odot}$, respectively. Moreover, they have used $T_1 = 11061 \; K$ for the primary component temperature which is known as $T_eff$ of the system in the Kepler eclipsing binary catalog (\citealt{Slawson2011}) and the temperature of the secondary component has been extracted $T_2 = 5671 \; K$ by using the ratio $\frac{T_2}{T_1}$ reported in this catalog. Furthermore, they have concluded that these temperatures for UZ Lyr were probably approximated less than the real value due to the position of this system on the H-R diagram (\citealt{Matson2017}). In the LAMOST-Kepler project, the effective temperature has been reported $T_eff = 11461 \pm 397 \;K$ with a spectral class A0IV (\citealt{Frasca2016}).\\
Periodic changes of orbital period of the UZ Lyr was investigated by \cite{Rafert1982} for the first time and a sinusoidal ephemeris was proposed for it. \cite{Hoffman2006} has studied the O-C curve of the eclipse minima times of this system, but he was unable to confirm the assumption of the existence of third body due to the high data scattering. However, after publication of the Kepler project data, a large number of eclipse minima times were available for this system. \cite{Gies2012} calculated the value $\frac{\dot{P}}{P}=2.29\times 10^{-6} \; year^{-1}$ based on this new data. Although they could not declare decisively about the third body existence, \cite{Gies2015} have re-examined the O-C curve of the UZ Lyr by using the new data and found that periodic changes are most likely due to the third body or the star spots movement. \cite{Borkovits2016} have studied the O-C curve and suggested the third body with a period of about $15 \; years$ and a minimum mass of $0.18 \; M_{\odot}$ for UZ Lyr.\\
The second system, BR Cyg, was introduced as an eclipsing binary by Leiner (1924) for the first time (\citealt{Wehinger1968}). In a study conducted by \cite{Kukarkin1958}, the spectral type of the primary component was found to be B9. The first photoelectric light curves of BR Cyg were obtained in two B \& V Johnson filters and then analyzed by \cite{Wehinger1968}. \cite{Kukarkin1971} re-studied on BR Cyg system and suggested spectral types of A5V and F0V for primary and secondary components of this system respectively. \cite{Giuricin1981} analyzed the Wehinger and Harmane photometric data and concluded that this spectral classification was incorrect probably. Also, they demonstrated that BR Cyg is a semi-detached system and fills the secondary component of its Roche lobe. \cite{Terrell2005} performed the CCD photometry of this system in four B, V, Rc, and Ic filters and analyzed its light curves. They selected $T_1 = 8900 \; K$ based on the B-V color index and obtained $q = 0.532$ for mass ratio. \cite{Liakos2009} investigated the possibility of pulsation, but they could not find any evidence of pulsation. In 2009, BR Cyg was included in the catalog of variable stars of Kepler project (\citealt{Pigulski2009}), and also in the Kepler eclipsing binary catalog in 2011 (\citealt{Prsa2011}). In the latest edition of this catalog, the effective temperature and the temperature ratio of this system have been reported as $T_eff = 8608 \; K$ and $\frac{T_2}{T_1}= 0.12$, respectively (\citealt{Kirk2016}). However, \cite{Armstrong2013} has informed the temperatures of $11056 \pm 603 \; K$ and $6278 \pm 1228 \; K$ for the primary and secondary components, respectively and also, the effective temperature has been obtained $10849 \; K$ based on the spectroscopy performed in the LAMOST-Kepler project (\citealt{Frasca2016}). \cite{Matson2017} have acquired $q = 0.516$ by applying the radial velocity curve of BR Cyg and calculated $M_1 = 3 \; M_{\odot}$ and $M_2 = 1.55 \; M_{\odot}$ for the mass of primary and secondary components.\\
\cite{Harmanec1973} analyzed the orbital period changes for the first time, but no regular behavior was observed in the O-C curve due to the high data scattering. After publication of the Kepler project data and obtaining a large number of minima data for this system, \cite{Gies2012} have examined the orbital period changes and reported the value of $\frac{\dot{P}}{P}=0.19 \times 10^{-6} \; year^{-1}$. They attributed the reason of these changes to either the movement of star spots or systematic changes in seasons . Moreover, they did not observe any sign of the existence of a third body for BR Cyg. These results are confirmed by \cite{Gies2015} and \cite{Zasche2015} too.
\cite{Zhang2019} combined the Kepler eclipsing binary catalog with the LAMOST data (for 1320 binaries) and found some parameters of this systems (e.g. period, $T_{eff}$, $\log{g}$, $M_1$, $R_1$, ...) but they did not analyse light curves of these systems. They assumed that the provided LAMOST spectra for each system, depended on the brighter component. Therefore, the atmosphere parameters measured from this spectra were obtained by using single star model.

\section{Period study}
The minima times of both UZ Lyr and BR Cyg are collected from the O-C gateway\footnote{http://var.astro.cz/ocgate/}, AAVSO\footnote{www.aavso.org} database and also, all minima times are extracted from the Kepler eclipsing binary stars database\footnote{http://keplerebs.villanova.edu}. In total, 968 primary eclipse minima times including 292 visual (vis), 13 photographic (pg) and 663 CCD or photoelectric (PE) minima times have been achieved for the UZ Lyr. Furthermore, 1345 MinI, including 268 vis, 9 pg, and 1068 CCD or PE have been obtained for the BR Cyg. This minima times have various validity, because they determined with different types of observations and by using different instruments. Also, some authors maybe used different numerical methods for extracting minima times from observational data. Therefore, we choose weight 1, 3 and 10 for whole group of observations obtained by vis, pg and pe or CCD techniques, respectively. This weighting method is used by many authors (e.g. \citealt{Zasche2008}; \citealt{Hanna2013}; \citealt{Ulas2020}).\\
{\bf UZ Lyr system:} The epoch and O-C value of this system are determined by using the linear ephemeris as follows (\citealt{Kirk2016}):

\begin{equation}
  (MinI)_{HJD}=2459954.335595+1.8912721\times E.
\label{eq1}
\end{equation}

\begin{figure}
\centering
\includegraphics[width=\textwidth, angle=0]{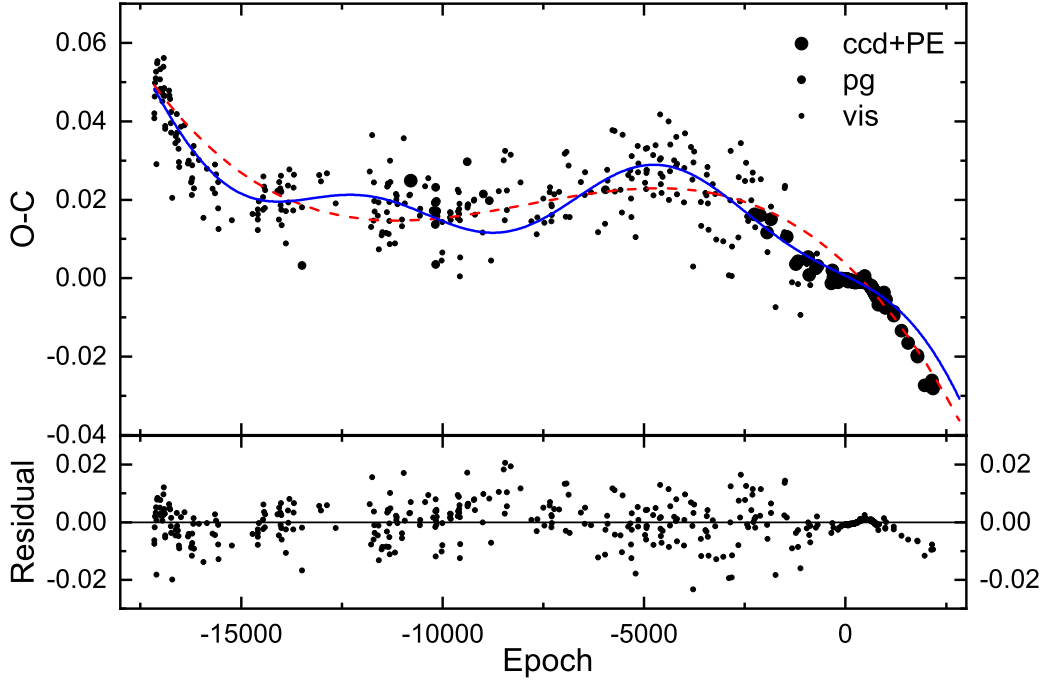}
\caption{The O-C curve of the UZ Lyr system. The dashed line shows fourth body and linear fit, and the solid line indicates the LTEs of both third and fourth body and linear function. The final residuals after removing the linear function, third and fourth body LTEs are shown in bottom.}
\label{fig1}
\end{figure}

The O-C curve is shown in Figure \ref{fig1}.

we used the linear, parabola and cubic function beside one or two light-time effects (LTE) for the O-C curve of this system. we concluded that a linear function with two LTEs is the best choose for this diagram, after comparing final fitting residuals. The coefficients of linear function determined by using the least squares method as follows:

\begin{equation}
O-C=-0.0486282(66)-6.646(25) \times 10^{-6} \times E.
\label{eq2}
\end{equation}

By using the slope of this linear function, the corrected orbital period of the binary system can be found as

\begin{equation}
P_{new}=1.89126545(25) \; day.
\label{eq3}
\end{equation}

After removing of the linear function from the O-C curve, the periodic behavior is apparent in the residuals in which these changes may be due to the LTE and existence of one or more components in the system. To analysis of the LTE in the O-C curve, the following equation has been used (\citealt{Irwin1959}):

\begin{equation}
(O-C)_{LTE}=\frac{K}{1-e^2\cos^2\omega}\times[\frac{1-e^2}{1-\cos\nu}\sin(\nu+\omega)+e\sin\omega],
\label{eq5}
\end{equation}

where $K$, $e$, $\omega$, and $\nu$ are the amplitude of changes, eccentricity, the longitude of periastron, and the true anomaly of the third or fourth body, respectively. We use Period04 Program (\citealt{Lenz2005}) to determine the initial values of orbital period of the third and fourth bodies. Orbital and physical parameters obtained by fitting Eq.\ref{eq5} to the O-C residuals are shown in Table \ref{tab1}. Figure \ref{fig1} illustrates the final simulated curve and residuals. After removing the LTEs of the third and fourth bodies and the linear function, it is obvious that the final residuals are randomly scattered around the zero line with no systematic behavior.

\begin{table}
\begin{center}
\caption[]{Orbital and physical parameters of the third and fourth bodies in the UZ Lyr.}
\label{tab1}

\begin{threeparttable}
 \begin{tabular}{ccc}
  \hline\noalign{\smallskip}
Parameter &  This work & \cite{Borkovits2016}\tnote{1}\\
  \hline\noalign{\smallskip}
$P_{3}(year)$  & 23.140(20) & 15.14\\ 
$e_{3}$  & 0.06(2) & 0.46(3)\\
$\omega_{3}(^\circ)$  & 63(2) & 242(9)\\
$K_{3}(day)$  & 0.00523(39) & -\\
$(T_0)_{3}(HJD)$  & 2447473(2524) & 2455955(153)\\
$a_{12}\sin i (AU)$  & 13.645(17) & 0.637(32)\\
$f(M_3)$ & 0.00138751(17) & 0.0011(2)\\
$M_{3,min} \; (M_{\odot})$ & 0.31494(13) & 0.17\\
$P_{4}(year)$  & 360(85) & -\\ 
$e_{4}$  & 0.60(11) & -\\
$\omega_{4}(^\circ)$  & 56(6) & -\\
$K_{4}(day)$  & 0.05011(69) & -\\
$(T_0)_{4}(HJD)$  & 2446644(874) & -\\
$a_{123}\sin i (AU)$  & 88.344(21) & -\\
$f(M_4)$ & 0.005990(9) & -\\
$M_{4,min} \; (M_{\odot})$ & 0.55187(29) & -\\
  \noalign{\smallskip}\hline
\end{tabular}
\begin{tablenotes}
    \item[1] \footnotesize{They used a cubic function and a LTE effect for the O-C curve of this system.}
\end{tablenotes}
\end{threeparttable}
\end{center}
\end{table}

{\bf BR Cyg system:} The epoch and O-C value of this system are determined by using the linear ephemeris (\citealt{Kirk2016}):

\begin{equation}
(MinI)_{HJD}=2454941.06399+1.3325545 \times E.
\label{eq6}
\end{equation}

\begin{figure}
\centering
\includegraphics[width=\textwidth, angle=0]{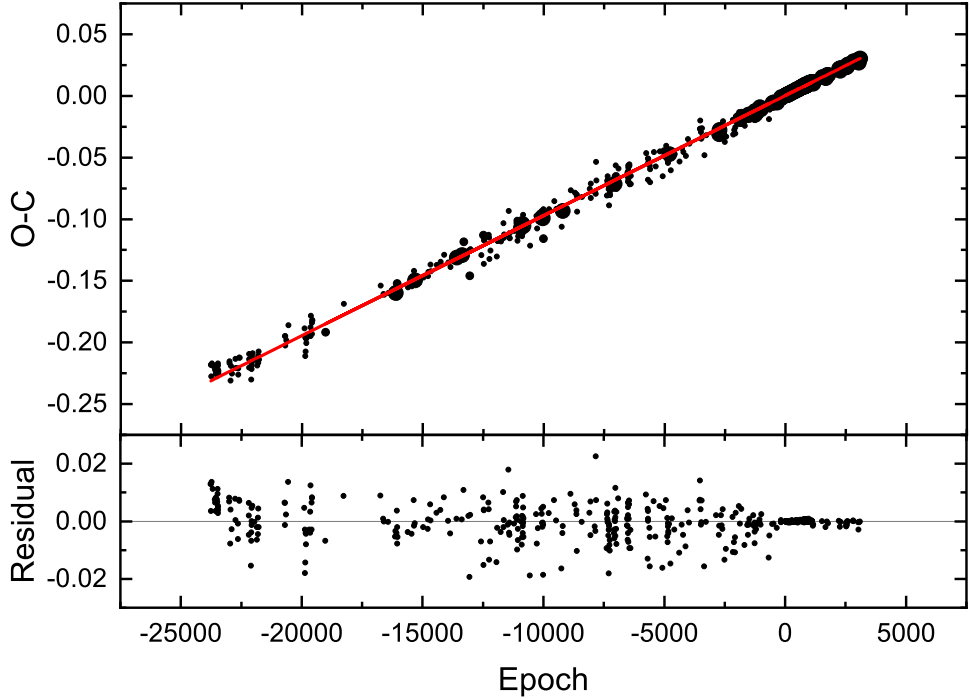}
\caption{The O-C curve of the BR Cyg. The solid line indicates the fit a linear function to the data points. The final residuals are shown in the lower part of the figure.}
\label{fig2}
\end{figure}

The obtained O-C curve is depicted in Figure \ref{fig2}.

At the first sight, the O-C curve resembles a linear function which its coefficients are determined by using the least squares method as follows:

\begin{equation}
O-C=3.55(33) \times 10^{-4}+9.737(13) \times 10^{-6} \times E.
\label{eq7}
\end{equation}

The linear changes of the O-C curve mean that the orbital period of the system has not been reported correctly in the initial ephemeris Eq.\ref{eq6}. With regard to the slope of the fitted line on the O-C curve, it is possible to obtain the following value as the new orbital period of this system:

\begin{equation}
P_{new}=1.332564237(13) \; day.
\label{eq8}
\end{equation}

After removing the linear function from the O-C curve, the final residuals are randomly scattered around the zero line with no systematic and periodic behaviors. Therefore, there is no evidence of the existence of a third body and also the mass transfer for this system. Owing to the high scattering of O-C data, effects with short periodicity and low amplitude are not observable in this diagram.

\section{Light curve analysis}
\begin{figure}
\centering
\includegraphics[width=\textwidth, angle=0]{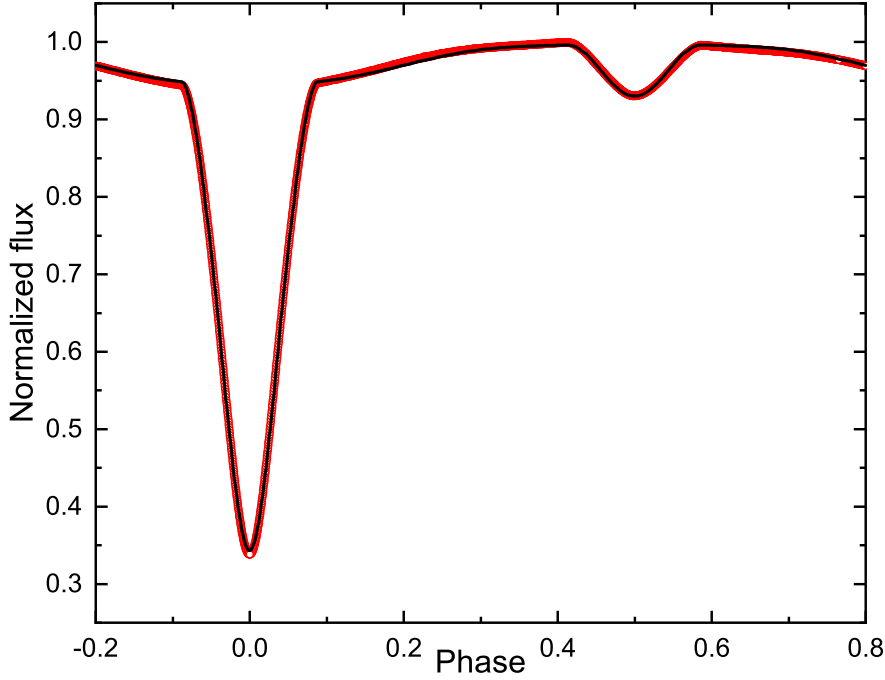}
\caption{Theoretical (solid black line) and observational (red dots) light curves of the UZ Lyr system.}
\label{fig3}
\end{figure}

\begin{figure}
\centering
\includegraphics[width=\textwidth, angle=0]{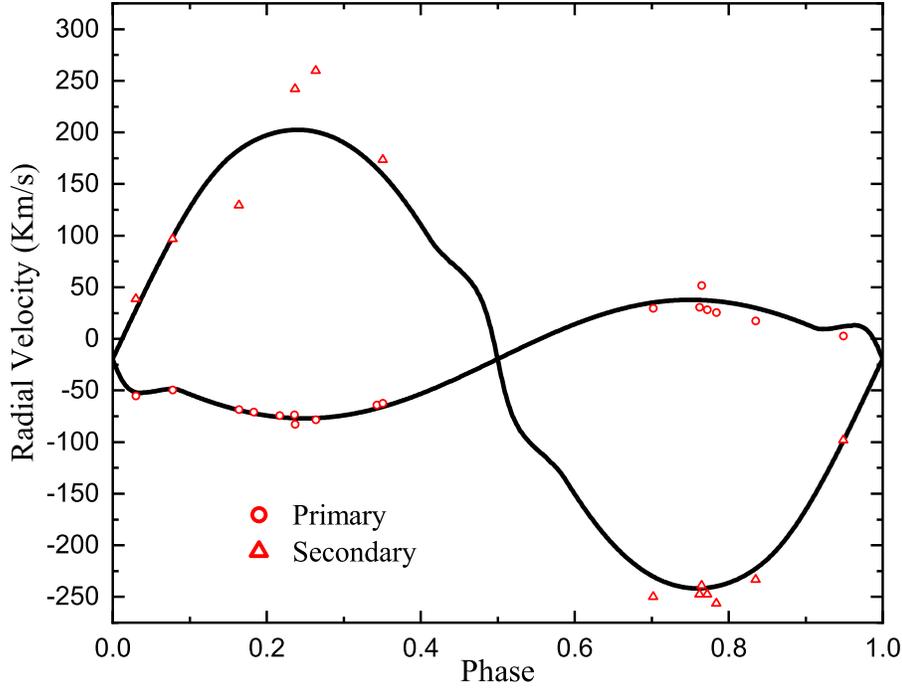}
\caption{Theoretical and observational radial velocity curves of the UZ Lyr.}
\label{fig4}
\end{figure}

The light curves data of both UZ Lyr and BR Cyg are extracted from the Kepler eclipsing binary stars database. We have used only SC (Short Cadence) data for plotting the light curves due to the high accuracy. Also, radial velocity data for these binary systems are adopted from (\citealt{Matson2017}) and we have used PHOEBE Legacy \footnote{http://phoebe-project.org} Program (\citealt{Prsa2005}). The light curves analysis are performed in "detached" mode firstly, but we could not find a good agreement between the theoretical and observational light curves. So the "semi-detached” mode are
chosen, with secondary component fills its Roche lobe, for both systems.\\
{\bf UZ Lyr:} Orbital phases were calculated by using the following linear ephemeris
\begin{equation}
(MinI)_{HJD}=2454941.06399+1.332564237 \times E,
\label{eq4}
\end{equation}
where the eclipse minima time chose from (\citealt{Kirk2016}) and the period obtained in this study. We use $T_1=11461 \; K$ (\citealt{Frasca2016}) for the temperature of the primary component as an constant parameter in the program. The initial mass ratio are chosen $q= 0.23$, which is based on the radial velocity data provided by (\citealt{Matson2017}) then this parameter defied as a free parameter in the program. Because $T_1>7200 \; K$, the values of $A_1 = 1$ (\citealt{Milne1926}) and $g_1 = 1$ (\citealt{Zeipel1924}) are used as constant parameters for the bolometric albedo and the gravity darkening coefficients, respectively. These two parameter are considered as free parameters for the secondary component. We apply logarithmic law for limb darkening coefficients which automatically calculated by the program (\citealt{Castelli2003}). Finally, Radial velocity and light curves data are analyzed simultaneously, and the results are given in Table \ref{tab2}. Figures \ref{fig3} and \ref{fig4} show the theoretical and observational light and radial velocity curves of the UZ Lyr, respectively.

\begin{figure}
\centering
\includegraphics[width=\textwidth, angle=0]{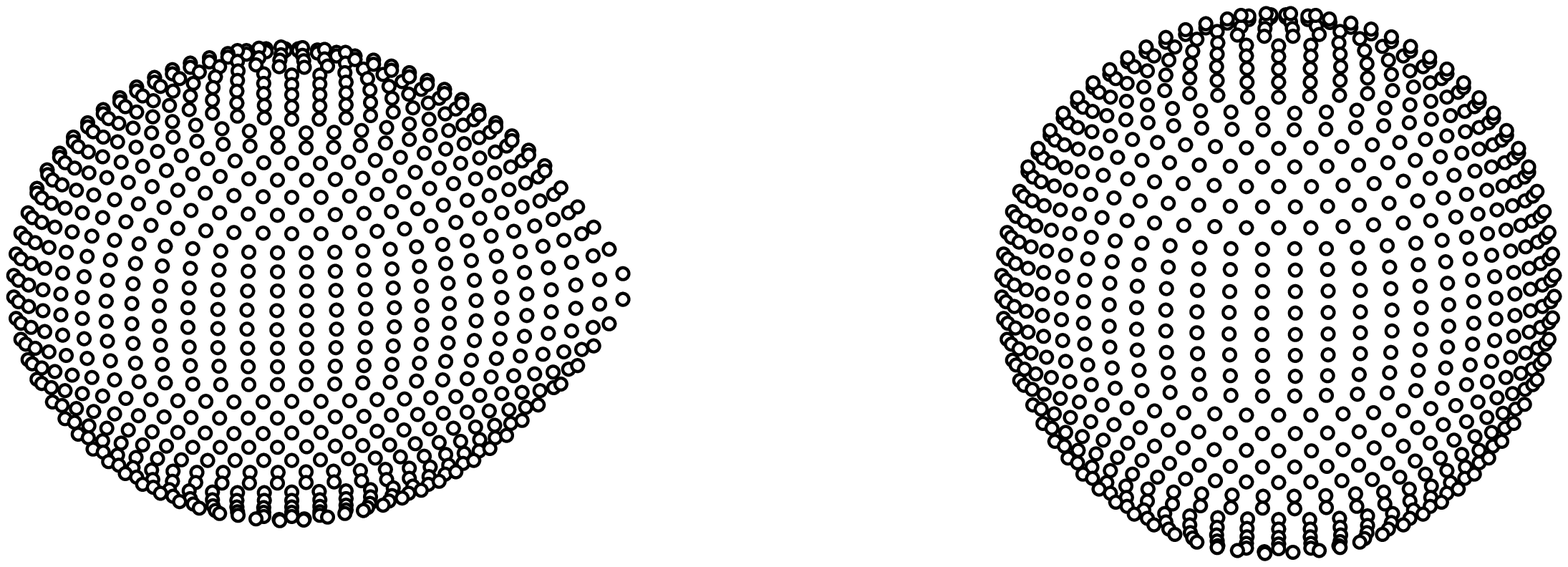}
\caption{Three-dimensional geometry of the UZ Lyr at phase 0.75. Secondary component fills its Roche lobe, the primary component retains a nearly spherical shape.}
\label{fig5}
\end{figure}

In Figure \ref{fig5} the three-dimensional geometry of system are displayed. Accordingly, the values of $K_1$ and $K_2$ are determined $57.4 \pm 4.5 \; \frac{km}{s}$ and $222 \pm 7 \; \frac{km}{s}$, respectively, which can be used for calculation of the absolute parameters of the UZ Lyr. Table \ref{tab3} shows absolute physical and geometrical parameters for this system.

\begin{figure}
\centering
\includegraphics[width=\textwidth, angle=0]{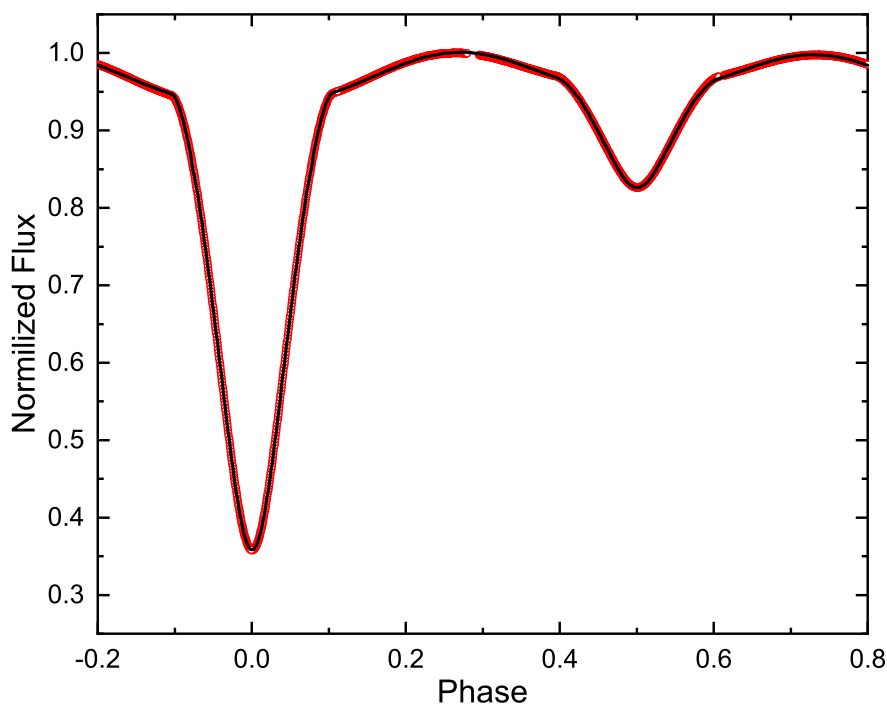}
\caption{Theoretical (solid black line) and observational (red dots) light curves of the BR Cyg.}
\label{fig6}
\end{figure}

\begin{figure}
\centering
\includegraphics[width=\textwidth, angle=0]{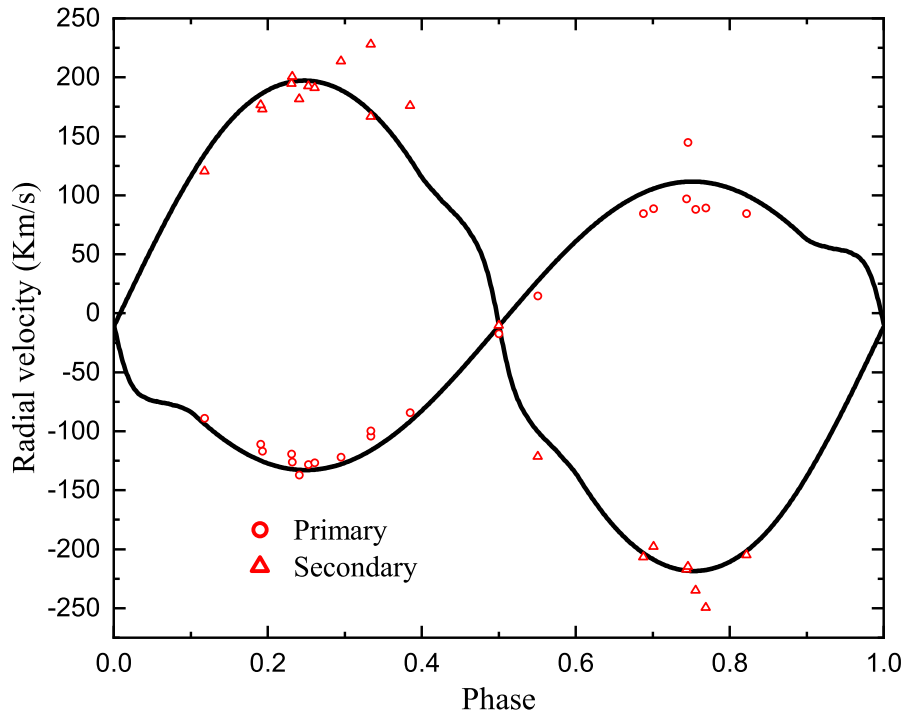}
\caption{Theoretical and observational radial velocity curves of the BR Cyg.}
\label{fig7}
\end{figure}

\begin{figure}
\centering
\includegraphics[width=\textwidth, angle=0]{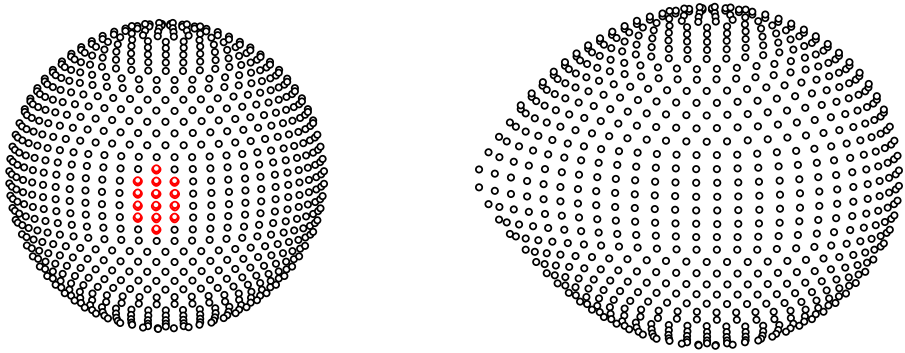}
\caption{Three-dimensional configuration of the BR Cyg at phase 0.25. Secondary component fills its Roche lobe, the primary component retains a nearly spherical shape. The position of a cold spot on the secondary star also has been shown.}
\label{fig8}
\end{figure}

{\bf BR Cyg:} The following linear ephemeris were used

\begin{equation}
(MinI)_{HJD}=2454941.06399+1.332564237 \times E,
\label{eq9}
\end{equation}

where the obtained period in this study and the eclipse minima time reported in (\citealt{Kirk2016}) are used. $T_1=10849 \; K$ has chosen based on spectroscopy performed in the LAMOST-Kepler project (\citealt{Frasca2016}). The mass ratio was considered as the free parameter with the initial value of $0.516$ based on the radial velocity curve presented by (\citealt{Matson2017}). Because the primary temperature is $T_1>7200 \; K$, the bolometric albedo, gravity and Limb darkening coefficients of the primary and secondary components, have been treated as the similar above section. The asymmetry at the light curve at the phases 0.25 and 0.75 can be attributed to the existence of cold or hot spots on one of the components (\citealt{OConnell1951}). So, we consider a cold spot on the secondary star and change its position, temperature ratio, and radius to achieve the best fit between the theoretical and observational light curves. Table \ref{tab4} gives the characteristics of this spot. Data of Radial velocity and light curves are analyzed simultaneously, and the results are given in Table \ref{tab2}. Figures \ref{fig6} and \ref{fig7} show the theoretical and observational light and radial velocity curves of the BR Cyg, respectively. Figure \ref{fig8} shows the three dimensional geometry of this system with the position of a cold spot on the secondary component.\\
Accordingly, the values of $K_1$ and $K_2$ are determined $122.26 \pm 3.1 \; \frac{km}{s}$ and $207.86 \pm 4.2 \; \frac{km}{s}$, respectively. The absolute physical parameters of the BR Cyg are calculated and given in Table \ref{tab3}. 

\begin{table}
\begin{center}
\caption[]{Results from simultaneous analysis of the radial velocity and light curves of the both UZ Lyr and BR Cyg systems.}
\label{tab2}

\begin{threeparttable}
 \begin{tabular}{ccccc}
  \hline\noalign{\smallskip}
Parameter &  UZ Lyr & \multicolumn{3}{c}{BR Cyg}\\
\cline{3-5}
 &  & This Work & \cite{Terrell2005}\tnote{1,2} & \cite{Giuricin1981}\tnote{1,3}\\
  \hline\noalign{\smallskip}
$i(^\circ)$  & 81.485(8) & 81.118(4) & 81.87(0.04) & 81.9(5)\\ 
$q=(\frac{M_2}{M_1})$  & 0.2242(3) & 0.5648(5) & 0.532(3) & 0.4\\
$\Omega_1$  & 3.851(2) & 4.118(1) & 3.872(8) & -\\
$T_1(K)$  & 11461 & 10849 & 8900 & 8800\\
$T_2(K)$  & 4259(7) & 6047(1.4) & 5698(5) & 5530(50)\\
$\frac{L_1}{L_1+L_2+L_3}$ & 0.97(3) & 0.798(3) & 0.905(1) & 0.844\\
$A_1$ & 1 & 1 & 1 & 1\\
$A_2$ & 0.59(1) & 0.322(1) & 0.5 & 0.5\\
$g_1$ & 1 & 1 & 1 & 0.25\\
$g_2$ & 0.33(2) & 0.248(7) & 0.32 & 0.08\\
$r_1 (pole)$ & 0.27(7) & 0.28(8) & - & -\\
$r_1 (point)$ & 0.28(8) & 0.29(9) & - & -\\
$r_1 (side)$ & 0.28(8) & 0.28(8) & - & -\\
$r_1 (back)$ & 0.28(8) & 0.29(9) & - & -\\
$r_2 (pole)$ & 0.24(24) & 0.31(16) & - & -\\
$r_2 (point)$ & 0.35 & 0.44 & - & -\\
$r_2 (side)$ & 0.25(28) & 0.32(19) & - & -\\
$r_2 (back)$ & 0.28(48) & 0.35(29) & - & -\\
$\sqrt{\frac{\Sigma(O-C)^2}{N}}$ & 0.0025 & 0.0011 & - & -\\
  \noalign{\smallskip}\hline
\end{tabular}
\begin{tablenotes}
    \item[1] \footnotesize{They analysed only light curves and therefore $q$ obtained by $q$-search method.}
    \item[2] \footnotesize{$T_1$ estimated based on B-V color-index of the system.}
    \item[3] \footnotesize{$T_1$ estimated based on spectral class of the system (A3V).}
\end{tablenotes}
\end{threeparttable}
\end{center}
\end{table}

\begin{table}
\begin{center}
\caption[]{Absolute physical and geometrical parameters of UZ Lyr and BR Cyg systems.}
\label{tab3}

\begin{threeparttable}
 \begin{tabular}{ccccccc}
  \hline\noalign{\smallskip}
Parameter &  \multicolumn{2}{c}{UZ Lyr} & & \multicolumn{3}{c}{BR Cyg}\\
\cline{2-3}
\cline{5-7}
 &  This work & \cite{Matson2017}\tnote{1} & & This Work & \cite{Matson2017}\tnote{1} & \cite{Giuricin1981}\tnote{2}\\
  \hline\noalign{\smallskip}
$a \; (R_{\odot})$  & 10.56(57) & 11.0(3) & & 8.80(48) & 8.44(8) & 8.8\\ 
$v_{com}\; (\frac{km}{s})$  & -19(3) & -25(2) & & -10.7(2.8) & -15.9(9) & - \\
$M_1 \; (M_{\odot})$  & 3.52(12) & 4.05(30) & & 3.24(16) & 3.00(7) & 2.5\\
$M_2 \; (M_{\odot})$  & 0.91(18) & 0.92(7) & & 1.91(7) & 1.55(3) & 1\\
$R_1 \; (R_{\odot})$  & 1.74(53) & - & & 1.93(61) & - & 2.3(2)\\
$R_2 \; (R_{\odot})$  & 1.75(81) & - & & 2.41(93) & - & 2.4(2)\\
$L_1 \; (L_{\odot})$ & 47.5(28.9) & - & & 47.13(29.7) & - & 28.18(9) \\
$L_2 \; (L_{\odot})$ & 0.99(92) & - & & 7.06(5.4) & - & 4.78(10)\\
  \noalign{\smallskip}\hline
\end{tabular}
\begin{tablenotes}
    \item[1] \footnotesize{They obtained these values only by radial velocity analysis.}
    \item[2] \footnotesize{They obtained these values by using mass-luminosity relation and assuming the primary is a main sequence star. Because they have not the radial velocity data of the system.}
\end{tablenotes}
\end{threeparttable}
\end{center}
\end{table}

\begin{table}
\begin{center}
\caption[]{Characteristics of a spot on the secondary component of the BR Cyg.}
\label{tab4}

 \begin{tabular}{cccc}
  \hline\noalign{\smallskip}
Latitude (rad) & Longitude (rad) &  Radius (rad) & $\frac{T_{Spot}}{T_2}$\\
  \hline\noalign{\smallskip}
1.57(9)  & 4.62(11) & 0.21(5) & 0.95(7) \\ 
  \noalign{\smallskip}\hline
\end{tabular}
\end{center}
\end{table}

\begin{figure}
\centering
\includegraphics[width=\textwidth, angle=0]{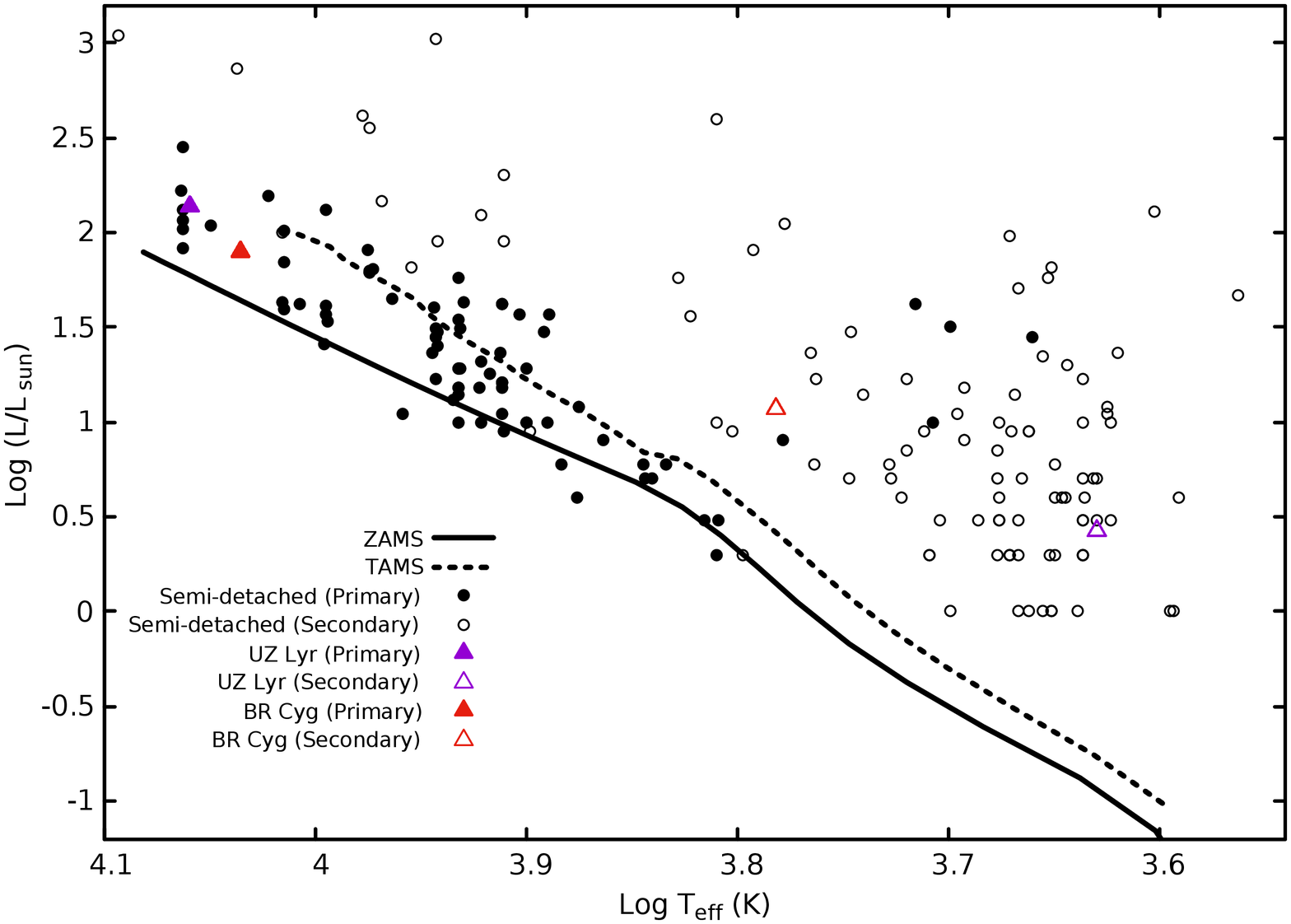}
\caption[Caption for LOF]{The position of primary and secondary components of UZ Lyr and BR Cyg in the H-R diagram. The examples of semi detached binary systems were taken from (\citealt{Malkov2020}). The ZAMS (zero age main sequence) and TAMS (terminal age main sequence) lines were obtained for the metal abundances like the Sun were used by MESA-Web\protect\footnotemark}
\label{fig9}
\end{figure}

\begin{figure}
\centering
\includegraphics[width=\textwidth, angle=0]{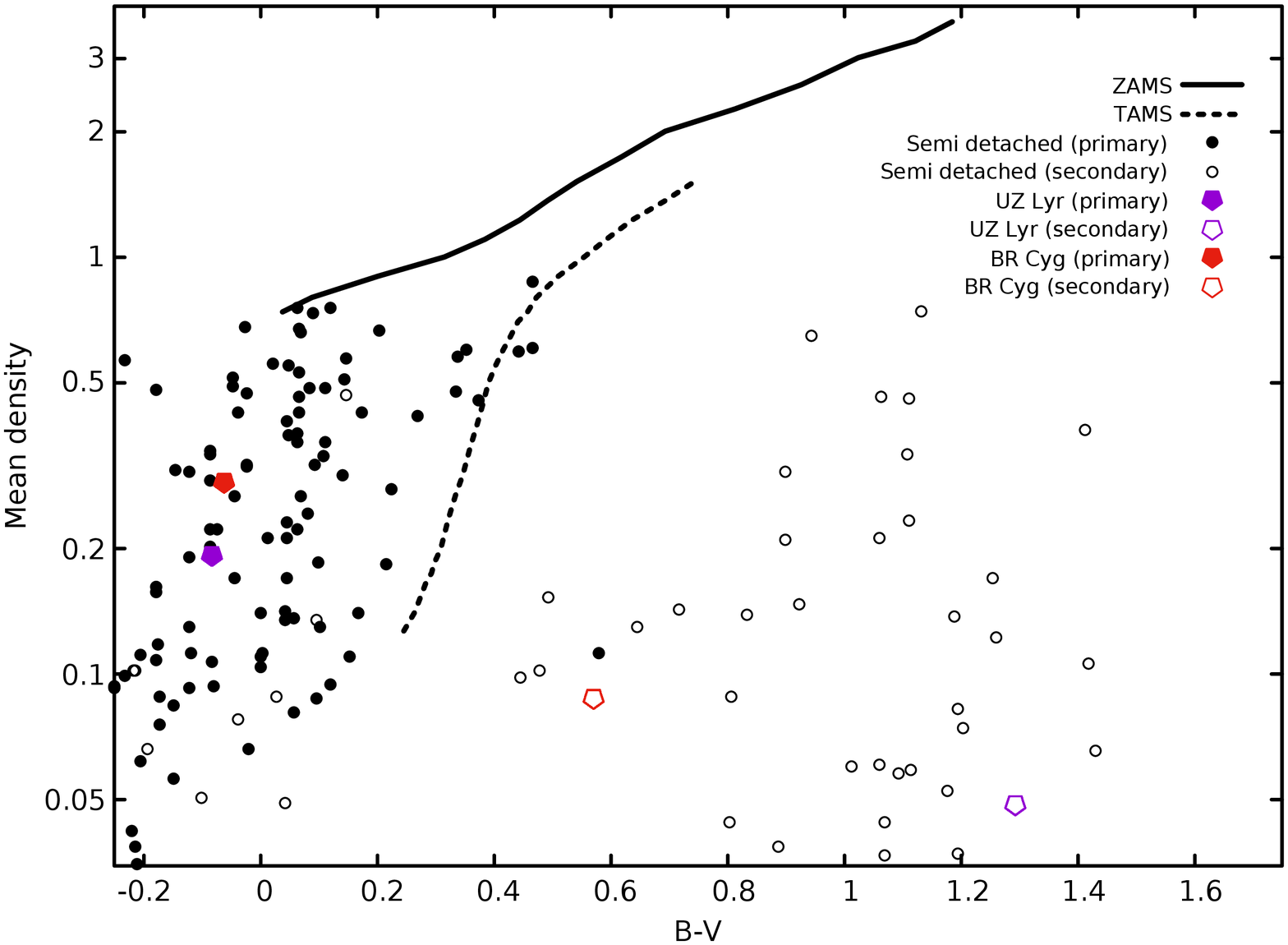}
\caption{Color index vs density diagram with the position of primary and secondary components of UZ Lyr and BR Cyg systems. The examples of semi detached binary systems were taken from (\citealt{Malkov2020}). The ZAMS and TAMS lines were adopted from (\citealt{Worthey2011}).}
\label{fig10}
\end{figure}

The H-R diagram is a useful tool for exploring the stars evolution position. Figure \ref{fig9} shows the position of the primary and secondary components of these systems in the H-R diagram based on the absolute parameters obtained for both the UZ Lyr and BR Cyg systems. For comparison, some other semi detached binaries (\citealt{Malkov2020}) are also shown in this figure. Accordingly, it is clear that in both systems, the secondary component left the main sequence toward the red giant phase, and the primary component is still remain on the main sequence. The color index vs density diagram also show the evolution status of stars (\citealt{Mochnacki1981, Mochnacki1984, Mochnacki1985}). Now by using the parameters of the Table \ref{tab3}, we can calculate the average density of the primary components of the UZ Lyr and BR Cyg systems as $0.193 \; \frac{gr}{cm^3}$ and $0.289 \; \frac{gr}{cm^3}$, and also for the secondary components of them are $0.049 \; \frac{gr}{cm^3}$ and $0.088 \; \frac{gr}{cm^3}$, respectively. Tables published in (\citealt{Worthey2011}), can be used for calculation of the B-V color index for the primary and secondary components of these systems which result $-0.083$ and $1.29$ for the UZ Lyr and $-0.063$ and $0.57$ for the BR Cyg, respectively. Finally, the specified position of both systems in the color index vs density diagram are shown in Figure \ref{fig10}.

\footnotetext{http://mesa-web.asu.edu}

\section{Discussion and conclusion}

The radial velocity and light curves of the eclipsing binary UZ Lyr were analyzed simultaneously for the first time, and the orbital and absolute parameters of this system were obtained. Firstly, light curve analysis was performed in "detached" mode, but there was no good agreement between the theoretical and observational light curve in this mode. So, we applied the "semi-detached” mode and consequently secondary component filled its Roche lobe. Orbital period changes study indicated a linear and two periodic behavior. So, the mass transfer does not have significant effects on orbital period changes. Also, the third and fourth bodies with a minimum mass of $0.31494 \; M_{\odot}$ and $0.55187 \; M_{\odot}$ respectively, and LTEs which is caused these periodic changes. \cite{Borkovits2016} assumed a cubic function and a third body with minimum mass of $0.17 \; M_{\odot}$ for this system, but we examined this possibility and we found our result (linear function and two LTEs effect) have better matched with data. According to our analysis, the effect of the third body light on the light curve of UZ Lyr was less than 1\%, which can be ignored due to the calculation errors. Thus, the third and fourth bodies lights were not observable in the light curve. Since the minimum mass achieved for these two additional components were within the mass range of white dwarfs, $0.2 < M_{WD} < 1.2 \; M_{\odot}$ (\citealt{Kepler2007, Kilic2007}), the third and fourth bodies would be either a white dwarf or a brown dwarf. This periodic behavior of the O-C curve also can be induced by other factors, such as magnetic activities (Applegate effect), but due to the absence of any spot on both of the components, it will be unlikely. After elimination of the effects of mass transfer and the third body of the O-C curve, no systematic behavior was observed in the residuals, which could be due to the data scattering. Thus, it is impossible to state about other factors which influence the orbital period changes.\\
The radial velocity and light curves of the BR Cyg binary were also analyzed simultaneously for the first time and the orbital and absolute parameters of this binary system were achieved. Firstly, the "detached" mode was used for the analysis of the light curve, but due to the absence of good results, the "semi-detached" mode was chosen. The orbital period changes study of the BR Cyg revealed a linear behavior. Therefore, the period suggested by \cite{Zasche2015} (Eq.\ref{eq6}) is not correct, so the new orbital period was found to be $1.332564237 \; day$ after determining the slope of the linear fit function to the O-C curve. After modification of the orbital period and redrawing the O-C curve, the final data points were scattered randomly around the zero line and no systematic behavior was observed. Thus, it is impossible to discuss more about orbital period changes with such available data. The positions of the primary and secondary components of both UZ Lyr and BR Cyg systems were shown on the color index vs density and H-R diagrams. As seen from figures 9 and 10, the position of components of both systems are in agreement with the other well known semi-detached binary systems. So, the secondary components have left the main sequence position toward the red giant phase but the primary components are still on the main sequence position. According to \cite{Popper1980} for semi-detached binary classification, both of those systems belong to the "Hot semi-detached" class (because the primary component temperature is larger than 7500 K). These systems are often will evolve with scenario named "case A" and therefore, the mass transfer to less massive component or to the out of system will start when the components were in the main sequence. At the first, the more massive component will lose most of it masses in non-conservative process same as stellar winds and secondary component will obtain just a little of mass. As a result, a non-evolved primary component will gain mass and usually it has smaller size and also is hotter and more massive than secondary component, but an evolved secondary component is more luminous and will lost it mass (\citealt{De Greve1986}). So, our results are in agreement with above scenario and positions of both systems on H-R diagram and color-index vs density diagram are shown correctly.

\label{lastpage}


\begin{thebibliography}{99}

\bibitem[Armstrong et al. (2013)]{Armstrong2013} Armstrong, D. J., Gomez Maqueo Chew, Y., Faedi, F., Pollacco, D., 2013, MNRAS, 437, 4.

\bibitem[Borkovits et al. (2016)]{Borkovits2016} Borkovits, T., Hajdu, T., Sztakovics, J., Rappaport, S., Levine, A., Biro, I. B., Klagyivik, P., 2016, MNRAS, 455, 4.

\bibitem[Castelli \& Kurucz (2003)]{Castelli2003} Castelli, F., Kurucz, R. L., 2003, In: N. Piskunov and D. F. Gray, eds., 210th Symposium of the International Astronomical Union, New Grids of ATLAS9 Model Atmospheres.

\bibitem[De Greve (1986)]{De Greve1986}De Greve, J.P., 1986, Space Sci Rev, 43, 139.

\bibitem[Frasca et al. (2016)]{Frasca2016} Frasca, A., Molenda-Zakowicz, J., De Cat, P., Catanzaro, G., Fu , J., N., Ren, A. B. et al., 2016, A\&A, 594, 39.

\bibitem[Gies et al. (2015)]{Gies2015} Gies, D. R., Matson, R. A., Guo, Z., Lester, K. V., Orosz, J. A., Peters, G. J., 2015, Astronomical Journal, 150, 6.

\bibitem[Gies et al. (2012)]{Gies2012} Gies, D. R., Williams, S. J., Matson, R. A., Guo, Z., Thomas, S. M., Orosz, J. A., Peters, G. J., 2012, Astronomical Journal, 143, 6.

\bibitem[Giuricin \& Mardirossian (1981)]{Giuricin1981} Giuricin, G., Mardirossian, F., 1981, A\&A, 96, 409.

\bibitem[Hanna \& Amin (2013)]{Hanna2013} Hanna, M. A., Amin, S. M., 2013, Journal of the Korean astronomical society, 46, 151.

\bibitem[Harmanec et al. (1973)]{Harmanec1973} Harmanec, P., Koubsky, P., Horn, J., Havelka, J., 1973, Bulletin of the Astronomical Institute of Czechoslovakia, 24, 311.

\bibitem[Hoffman et al. (2006)]{Hoffman2006} Hoffman, D. I., Harrison, T. E., McNamara, B. J., Vestrand, W. T., Holtzman, J. A., Barker, T., 2006, The Astronomical Journal, 132, 6.

\bibitem[Irwin (1959)]{Irwin1959} Irwin, J. B., 1959, Astronomical Journal, 46, 149.

\bibitem[Kalimeris et al. (1994)]{Kalimeris1994} Kalimeris, A., Rovithis-Livaniou, H., Rovithis, P., 1994, A\&A, 282, 3.

\bibitem[Kepler et al. (2007)]{Kepler2007} Kepler, S. O., Kleinman, S. J., Nitta, A., Koester, D., Castanheira, B. G., Giovannini, O. et al., 2007, MNRAS, 375, 4.

\bibitem[Kilic et al. (2007)]{Kilic2007} Kilic, M., Prieto, C. A., Brown, W. R., Koester, D., 2007,The Astrophysical Journal, 660, 2.

\bibitem[Kirk et al. (2016)]{Kirk2016} Kirk, B., Conroy, K., Prsa, A., Abdul-Masih, M., Kochoska, A., Matijevic, G. et al., 2016, The Astronomical Journal, 151, 3.

\bibitem[Koch et al. (1979)]{Koch1979} Koch, R. H., Wood, F. B., Florkowski, D. R., Oliver, J. P., 1979, IBVS, 1709.

\bibitem[Kukarkin et al. (1958)]{Kukarkin1958} Kukarkin, B. V., Parenago, P. P., Efremov, Yu. I., Kholopov, P. N., 1958, General Catalog of Variable Stars, 1, 220.

\bibitem[Kukarkin et al. (1971)]{Kukarkin1971} Kukarkin, B. V., Kholopov, P. N., Efremov, Yu. N., Kukarkina, N. P., Kurochkin, N. E., Medvedeva, G. I., Perova, N. B., Pskovskij, Yu. P., Fedorovich, V. P., Frolov, M. S., 1971, First supplement to the third edition of the general catalogue of variable stars, p.41.

\bibitem[Lenz \& Breger (2005)]{Lenz2005} Lenz, P., Breger, M., 2005, Communications in Asteroseismology, 146.

\bibitem[Liakos \& Niarchos (2009)]{Liakos2009} Liakos, A., Niarchos, P., 2009, Communications in Asteroseismology, 160, 2.

\bibitem[Malkov (2020)]{Malkov2020} Malkov, O. Y., 2020, MNRAS, 491, 4.

\bibitem[Matson et al. (2017)]{Matson2017} Matson, R. A., Gies, D. R., Guo, Z., Williams, S. J., 2017,The Astronomical Journal, 154, 216.

\bibitem[Milne (1926)]{Milne1926} Milne, E. A., 1926, MNRAS, 87, 1.

\bibitem[Mochnacki (1981)]{Mochnacki1981} Mochnacki, S. W., 1981, The Astrophysical Journal, 245, 650.

\bibitem[Mochnacki (1984)]{Mochnacki1984} Mochnacki, S. W., 1984, The Astrophysical Journal Supplement Series, 55, 551.

\bibitem[Mochnacki (1985)]{Mochnacki1985} Mochnacki, S. W., 1985, The Astrophysical Journal Supplement Series, 59, 445.

\bibitem[Nijland (1931)]{Nijland1931} Nijland, A. A., 1931, Astronomische Nachrichten, 242, 5.

\bibitem[O'Connell (1951)]{OConnell1951} O’Connell, D. J. K., 1951, MNRAS, 111, 6.

\bibitem[Pigulski et al. (2009)]{Pigulski2009} Pigulski, A., Pojmański, G., Pilecki, B., Szczygiel, D. M., 2009, Acta Astronomica, 59, 1.

\bibitem[Popper (1980)]{Popper1980}Popper, D. M., 1980, ARA\& A, 18, 115.

\bibitem[Prsa et al. (2011)]{Prsa2011} Prsa, A., Batalha, N., Slawson, R. W., Doyle, L. R., Welsh, W. F., Orosz, J. A. et al., 2011, Astronomical Journal, 141, 3.

\bibitem[Prsa \& Zwitter (2005)]{Prsa2005} Prsa, A., Zwitter, T., 2005, The Astrophysical Journal, 628, 1.

\bibitem[Rafert (1982)]{Rafert1982} Rafert, J. B., 1982, PASP, 94, 559.

\bibitem[Slawson et al. (2011)]{Slawson2011} Slawson, R. W., Prsa, A., Welsh, W. F., Orosz, J. A., Rucker, M., Batalha, N. et al., 2011, Astronomical Journal, 142, 5.

\bibitem[Terrell \& Gross (2005)]{Terrell2005} Terrell, D., Gross, J., 2005, IBVS, 5646.

\bibitem[Ulas et al. (2020)]{Ulas2020} Ulas, B., Gazeas, K., Liakos, A., Ulusoy, C., Stateva, I., Erkan, N., Napetova, M., Iliev, I. K., 2020, Acta Astronomica, 70, 219.

\bibitem[Wehinger (1968)]{Wehinger1968} Wehinger, P. A., 1968, The Astronomical Journal, 73, 163.

\bibitem[Worthey \& Lee (2011)]{Worthey2011} Worthey, G., Lee, H., 2011, The Astrophysical Journal Supplement Series, 193, 1.

\bibitem[Zasche et al. (2008)]{Zasche2008} Zasche, P., Liakos, A., Wolf, M., Niarchos, P., 2008, New Astronomy, 13, 405.


\bibitem[Zasche et al. (2015)]{Zasche2015} Zasche, P., Wolf, M., Kucakova, H., Vrastil, J., Jurysek, J., Masek, M., Jelinek, M., 2015, Astronomical Journal, 149, 6.

\bibitem[Zeipel (1924)]{Zeipel1924} Zeipel, H. v., 1924, MNRAS, 84, 9.

\bibitem[Zhang et al. (2019)]{Zhang2019} Zhang, J., Qian, S. B., We, Y., Zhou, X., 2019, ApJS, 244, 43.

\end{thebibliography}
\end{document}